\documentclass[11pt]{article}

\usepackage[]{graphicx}

\title{Physical-resource demands for scalable quantum computation%
\footnote{To be published in the proceedings of the SPIE Conference
on Fluctuations and Noise in Photonics and Quantum Optics, Santa Fe,
New Mexico, June 1--4, 2003}}

\author{Carlton M.~Caves,$^{\hbox{\small{(a)}}}$%
\footnote{\tt caves@info.phys.unm.edu}\quad Ivan H.~Deutsch,%
$^{\hbox{\small{(a)}}}$%
\footnote{\tt ideutsch@info.phys.unm.edu}\quad and
Robin Blume-Kohout$^{\hbox{\small{(b)}}}$%
\footnote{\tt rbk@socrates.berkeley.edu}
\\ \hbox{} \\
$^{\hbox{\small{(a)}}}$Department of Physics and Astronomy,
University of New Mexico, \\
Albuquerque, NM~87131--1156, USA \\
$^{\hbox{\small{(b)}}}$Los Alamos National Laboratory, Mail Stop B210,
Los Alamos, NM~87545, USA
}

\begin{document}

\maketitle

\begin{abstract}
The primary resource for quantum computation is Hilbert-space
dimension.  Whereas Hilbert space itself is an abstract construction,
the number of dimensions available to a system is a physical quantity
that requires physical resources.  Avoiding a demand for an
exponential amount of these resources places a fundamental constraint
on the systems that are suitable for scalable quantum computation.
To be scalable, the number of degrees of freedom in the computer must
grow nearly linearly with the number of qubits in an equivalent
qubit-based quantum computer.
\end{abstract}

\section{Introduction}
\label{sect:intro}

Quantum computation is an alluring long-term goal for the emerging
field of quantum information science \cite{NSF,NielsenChuang}.  In
this paper we address the question of what physical resources are
required for quantum computation and, in particular, how the required
resources scale with problem size.  By determining how to avoid a
physical-resource demand that increases exponentially with problem
size, we establish necessary conditions for a physical system to be a
scalable quantum computer.

The initial step in a quantum computation is to store classical
information (the input) as some quantum state of the computer.  The
computer then runs through a carefully controlled sequence of unitary
operations and/or measurements (the program).  At the completion of
the computation, the answer (the output) is stored as classical
information that can be read out with high probability by a
measurement.  The power of a quantum computer lies somewhere in the
murky region between the classical input and the classical output---a
region where classical, realistic descriptions fail.

Ask for the crucial property of that murky region, and you will get
nearly as many answers as there are quantum information scientists:
the superposition principle of quantum mechanics and associated
quantum interference and quantum parallelism; quantum entanglement;
the use of entangling unitary operations; the collapse of the wave
function after measurement and associated information-disturbance
trade-offs.  All of these distinguish quantum systems from classical
ones.  How are we to decide which is the crucial quantum feature?

We argue that a quantum computer's power stems from the murky region
itself: a quantum computer escapes the bounds of classical
information processing because there is no efficient realistic
description of what happens between the classical input and the
classical output.  The ability to access arbitrary states in Hilbert
space is what leads to situations where there is no efficient
realistic desciption.

It is difficult to pin down the source of a quantum computer's power
because arbitrary states can be accessed in very different physical
systems---different hardware---using very different control
techniques---different software.  Yet no matter how a quantum
computation is packaged, we can identify one universal prerequisite:
the computer must have a Hilbert space large enough to accommodate
the computation.  If the computer is to be a general-purpose
computer, able in principle to solve problems of arbitrary size, it
must have a Hilbert space whose dimension is capable in principle of
growing exponentially with problem size.  Hilbert space is essential
for quantum computation, and the primary resource is {\em
Hilbert-space dimension}.

Hilbert spaces of the same dimension are {\em fungible}.  What can be
done in one can be done in principle in any other of the same
dimension: simply map one Hilbert space onto the other, including all
the subsystems, operations, and measurements.  Which Hilbert space is
used to represent and process quantum information becomes important
only when further physical considerations are introduced. Though
Hilbert spaces are fungible, the physical systems described by those
Hilbert spaces are not, because {\em we don't live in Hilbert space}.
A Hilbert space gets its connection to the world we live in through
the physical quantities---position, linear momentum, energy, angular
momentum---of the system that is described by that Hilbert space.
These physical quantities arise naturally from spacetime symmetries
and the system Hamiltonian.  They are the ``handles'' that permit us
to grab hold of a quantum system and manipulate it, and they are the
physical resources that must be supplied to access various parts of
the system Hilbert space.  The crucial {\em physical\/} question for
quantum computation is the following: {\em how much of these
resources is required to achieve a Hilbert-space dimension sufficient
for a computation?}  This is the question we address in this paper.

Quantum mechanics constrains our description of physical systems
sufficiently that we can formulate the question of physical-resource
demands in a general way.  We find that to avoid supplying an amount
of some physical resource that grows exponentially with problem size,
the computer must be made up of subsystems---degrees of freedom in
the simple analysis presented here---whose number grows nearly
linearly with the number of qubits required in an equivalent quantum
computer. This thus becomes a fundamental requirement for a system to
be a {\em scalable\/} quantum computer.\footnote{Since in this paper
we are interested in comparing how different systems use physical
resources, we use the term {\em quantum computer\/} for any physical
system that has the required Hilbert-space dimension, and we reserve
the term {\em scalable quantum computer\/} for systems that can
provide the required Hilbert-space dimension efficiently.}

We emphasize that this requirement is an initial barrier that must be
surmounted by proposals for scalable quantum computation, before such
proposals confront the difficult and necessary tasks of
initialization, control, protection from errors, and readout.
Surmounting this initial barrier does not guarantee that a proposal
can meet the further requirements; it is a necessary, but by no means
sufficient requirement for a scalable quantum computer.  The point of
this paper is that one can draw general conclusions about the
physical systems that can be used for quantum computation just by
considering whether the system can efficiently provide the {\em
primary\/} resource of Hilbert-space dimension, without getting
enmeshed in questions about the other necessary requirements for the
operation of a quantum computer.

The present paper is an abbreviated version of a more extensive
analysis published elsewhere \cite{original}.  Here we restrict our
consideration to systems of particles that can be described by
ordinary quantum mechanics.  For these systems, the subsystems can be
identified with the {\em degrees of freedom\/} of the particles.
Though this simple degrees-of-freedom analysis contains the essence
of our conclusions---indeed, the essence is summarized succinctly in
Fig.~\ref{fig1}---the original paper extended the analysis to more
general systems that require a description in terms of quantum
fields.  The reader interested in this more general analysis and in a
more extensive list of references is urged to consult the original
paper \cite{original}.

\section{Degrees-of-freedom analysis of resource requirements}
\label{sec:dofanalysis}

\subsection{Role of Planck's constant}
\label{sec:Planck}

Dimensionless quantities in physics are determined by writing the
relevant physical quantities in terms of a relevant scale.  For the
dimension of a system's Hilbert space, Planck's constant $h$ sets the
scale; the available number of Hilbert-space dimensions is determined
by writing an appropriate combination of physical quantities, the
{\em action}, in units of $h$.

The analysis of resource demands is particularly simple for systems
of particles described by ordinary quantum mechanics, for which the
subsystems can be identified with the degrees of freedom of the
particles.  The quantum state of such a computer is described in a
Hilbert space that is a tensor product of the Hilbert spaces of the
degrees of freedom.

A degree of freedom corresponds to a pair of (generalized) canonical
co\"ordinates, position $q$ and momentum $p$.  The physical resources
are the ranges of positions and momenta, $\Delta q$ and $\Delta p$,
used by the computation.  The physically relevant measure of these
resources is the corresponding phase-space area or {\em action},
$A=\Delta q\Delta p$.  For a degree of freedom that is an intrinsic
angular momentum $J$, we can use $\Delta q=2\pi$ and $\Delta p=\Delta
J$, thus giving $A=2\pi\Delta J$.  The connection to Hilbert space
comes from the fact that a quantum state occupies an area in phase
space given by Planck's constant $h\,$; orthogonal states correspond
roughly to nonoverlapping areas, each of area $h$.  Thus the
available dimension of the Hilbert space for a single degree of
freedom is given approximately by $A/h$.  The goal of scalability is
to avoid having to supply an action resource $A$ for any degree of
freedom that grows exponentially with problem size.

\subsection{Degrees-of-freedom analysis}
\label{sec:dof}

We measure the Hilbert-space dimension required for a quantum
computation in qubit units: let the problem size for a computation be
$n$, and let $N={\bf F}(n)$ be the number of qubits required for the
computation, assuming an optimal qubit algorithm that requires only a
polynomial number of qubits, an example being Shor's factoring
algorithm \cite{NielsenChuang}.  Here and below, bold type denotes a
function that is bounded above by a polynomial.  The Hilbert-space
dimension needed for the computation is $2^N=2^{\mbox{\scriptsize\bf
F}(n)}$. Using qubit units, we see that the required Hilbert-space
dimension grows exponentially with problem size.  We assume that there
is no more efficient algorithm in a Hilbert space with some other
structure than the qubit tensor-product structure, this being part of
our assumption that Hilbert spaces are fungible.

Suppose now that the $j$th degree of freedom supplies an action
$A_j$.  The Hilbert space of the entire system is a tensor product of
the Hilbert spaces for the degrees of freedom, so the overall Hilbert
space has dimension
\begin{equation}
2^N\sim{A_1\over h}\cdots{A_T\over h}={V\over h^T}\;,
\end{equation}
where $V$ is the phase-space volume used by the computation. If $T$
grows more slowly than linearly with $N$ (within specific logarithmic
corrections discussed below), at least one of the actions must grow
exponentially with $N$, thus requiring an exponential amount of some
physical resource.  In contrast, if $T$ grows linearly with $N$, no
degree of freedom has to supply an increasing amount of action, which
makes the system a candidate for a scalable quantum computer.

\begin{figure}
\begin{center}
\includegraphics[width=5.5in]{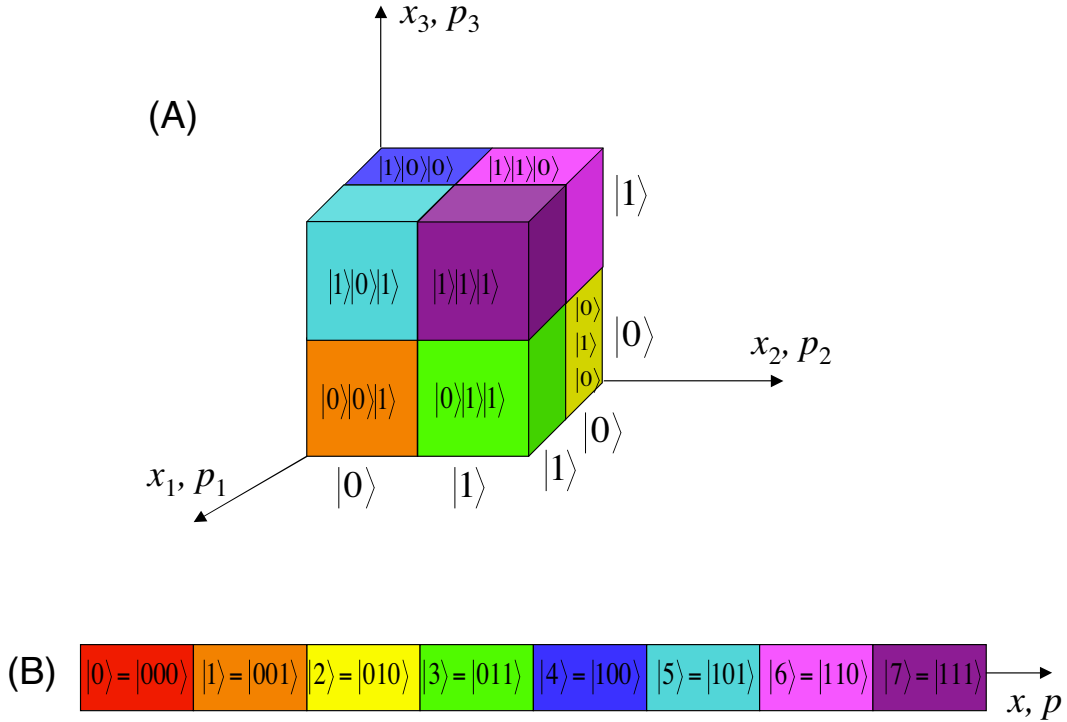}
\end{center}
\vspace{-6pt}
\caption{\label{fig1}\small
{\bf Using many degrees of freedom to save resources.} Orthogonal
basis states for an eight-dimensional Hilbert space are depicted
schematically as nonoverlapping phase-space cells in the phase space
of three degrees of freedom (A), each of which uses an action $\sim
2h$, or in the phase space of a single degree of freedom (B). Phase
space is pictured at half its actual dimension by letting the axes
represent both the position and momentum co\"ordinates for a degree
of freedom; one can think of the axes as measuring the amount of
action used by a degree of freedom. To accommodate the eight states,
the single degree of freedom requires three times as much action as
does each of the three degrees of freedom.  If one adds degrees of
freedom to (A), the phase-space volume---and hence the Hilbert-space
dimension---doubles as each degree of freedom is added and thus grows
exponentially with the number of degrees of freedom, whereas the
physical resources grow linearly with the number of degrees of
freedom and thus logarithmically with the Hilbert-space dimension.
The result is a scalable resource requirement.  In contrast, for the
single degree of freedom in (B), the required resources grow linearly
with phase-space volume and Hilbert-space dimension; to achieve the
same Hilbert-space dimension as for the scalable case requires
physical resources that are exponentially larger.  The use of many
degrees of freedom, with the number of degrees of freedom growing
nearly linearly with the corresponding number of qubits, allows the
required phase-space volume to be folded up into a hypercube so that
no degree of freedom has to provide an exponential amount of action.
As shown, the basis states for both situations can be labeled either
by unary or binary numbers, this being an example of the fungibility
of Hilbert spaces. The labeling, however, cannot alter the physics:
the single degree of freedom is a {\em physically\/} unary
realization of the Hilbert space, which uses exponential resources
asymptotically, whereas the multiple degrees of freedom in (A)
provide a {\em physically\/} binary realization of the same Hilbert
space, which uses resources efficiently.}
\end{figure}

It is useful to summarize this simple result, as it is the foundation
for all our further conclusions.  The physical resources are the
quantities that label the axes of a (generalized) phase space that
has two axes for each degree of freedom.  The number of Hilbert-space
dimensions available for a computation is proportional to the total
phase-space volume.  If the number of degrees of freedom grows
linearly in $N$, the phase-space volume needed to accommodate the
Hilbert-space dimension can be folded up into a hypercube in phase
space without requiring an exponentially increasing contribution
along any direction in phase space.  In contrast, if the number of
degrees of freedom grows more slowly than linearly in $N$ (within the
logarithmic corrections discussed below), some phase-space direction
must supply an exponentially increasing amount of the corresponding
physical resource.  This simple argument is depicted schematically in
Fig.~\ref{fig1}.

To formulate a more precise statement, we specialize to the case of
$T$ identical degrees of freedom, each of which supplies an action
$A$. In this situation, the total number of Hilbert-space dimensions
satisfies $(A/h)^T\sim2^N$, which gives
\begin{equation}
A/h\sim2^{N/T}\;.
\end{equation}
In order to avoid an exponential resource demand, $A/h$ must grow
polynomially with $N$,\footnote{We follow the computer-science
convention of referring to any superpolynomial growth as
exponential.} which means that the number of degrees of freedom
increases as\footnote{We use base-2 logarithms.}
\begin{equation}
T\sim{N\over\log{\bf P}(N)}\;,
\label{eq:quasilinear}
\end{equation}
where ${\bf P}(N)$ is a function bounded above by a polynomial.  We
say that $T$ grows {\em quasilinearly\/} with $N$ and that the system
is {\em scalable}, having a {\em scalable tensor-product structure}.

It is instructive to distinguish three cases:
\begin{enumerate}
\begin{item}
$T$ grows more slowly than linearly with $N$.  If $T$ grows
quasilinearly, as in Eq.~(\ref{eq:quasilinear}), then $A/h\sim{\bf
P}(N)$, and the system is scalable.  If $T$ grows more slowly than
quasilinearly with $N$, $A/h$ grows exponentially with $N$, leading
to an exponential demand for physical resources.
\end{item}
\begin{item}
$T$ grows faster than linearly with $N$.  In this case, $A/h$ goes to
one as $N$ increases, implying that each degree of freedom asymptotes
to a one-dimensional Hilbert space.  This means that the present
analysis in terms of independent degrees of freedom breaks down and
should be replaced by a counting of the excitations of a quantum
field, which properly takes into account the resources used by
unoccupied field modes \cite{original}.
\end{item}
\begin{item}
$T=N/\log D$ grows strictly linearly with $N$.  For $D<2$, the
present analysis breaks down, and we again need the analysis of
quantum fields to reach a sensible conclusion \cite{original}.  For
$D\ge2$, each degree of freedom is a $D$-level system, i.e., a {\em
qudit\/} instead of a qubit.  Though this is the special case of
quasilinear growth in which ${\bf P}(N)=D$, we consider it separately
because it is the most important scalable case, in that the action
supplied by each system, $A/h\sim D$, is independent of problem size.
Scaling is achieved simply by adding degrees of freedom, without
having to change the Hilbert-space dimension supplied by each degree
of freedom.  We say that this kind of system is {\em strictly
scalable\/} and has a {\em strictly scalable tensor-product
structure}.  Most quantum computing proposals are of this sort.
\end{item}
\end{enumerate}

Had we focused on the total action resource,
\begin{equation}
TA/h\sim T2^{N/T}\;,
\label{eq:totalaction}
\end{equation}
instead of on the action resource per degree of freedom, we would
have reached the same conclusions regarding scaling.  For a scalable
system, the total action resource grows as $TA/h\sim N{\bf
P}(N)/\log{\bf P}(N)$; only for strictly scalable systems is the
total action resource linear in $N$.

\subsection{Quantum computing in a single atom}
\label{sec:singleatom}

An illuminating extreme example of the nonscalable systems in case~1
is the attempt to implement quantum computing in a single
atom \cite{Ahn}, single molecule with a fixed number of
atoms \cite{Zadoyan,Lozovoy}, or large spin \cite{Leuenberger}.
Advances in laser spectroscopy with ultrashort pulses have allowed
researchers to manipulate and measure the electronic wave function in
an atom \cite{Noel} or both electronic and rotational/vibrational wave
functions in a molecule \cite{Rabitz} with exquisite precision. It is
natural to wonder whether these tools for coherent control of quantum
states can be applied to quantum computing.

For illustration, consider the simplest hypothetical model, quantum
computing in a hydrogen atom.  Characteristic atomic units of length,
momentum, and energy are formed from the physically important
constants: the electron charge and mass, $e$ and $m$, and the quantum
of action, $\hbar$.  If we ignore spin, Bohr's formula for quantizing
the action gives the familiar expressions for the energy, radius, and
momentum of a stationary state with principle quantum number $n$,
\begin{equation}
E_n=-{1\over {2n^2}}\,{{e^2} \over {a_0}}\;,\quad
r_n=n^2a_0\;,\quad
p_n={1 \over n}\,{\hbar  \over {a_0}}\;,
\end{equation}
where $a_0 = \hbar^2 / m e^2$ is the Bohr radius.  The dimension of
the Hilbert space spanned by all bound states from the ground state
up to a maximum principle quantum number $n$ is
\begin{equation}
\sum_{k=1}^n \sum_{l=0}^{k-1}
(2l+1)\sim {1\over 3}n^3 \sim \left( {r_np_n \over \hbar } \right)^3\;.
\end{equation}
The final expression has just the form we expect.  Without spin the
internal states of the hydrogen atom have three degrees of freedom,
signaled by the 3 in the exponent and corresponding to the three
co\"ordinates of relative motion of the electron and proton. Each
degree of freedom is allotted an action $A\sim r_n p_n$, which
provides enough phase space for $\sim A/h$ orthogonal states in
Hilbert space.

Demanding that the atomic Hilbert space have a dimension $2^N$
requires that the radial co\"or\-di\-nate scale as $r_n\sim
2^{2N/3}a_0$. The exponential growth of this co\"or\-di\-nate with
problem size implies that quantum control in a single atom {\em
cannot\/} be used for scalable quantum computation.  For instance, to
implement a quantum computation requiring $N=100$ qubits, the atomic
radius must be $r_n\sim 10^{20}a_0=6\times 10^6\,$km, about 5 times
the diameter of the Sun.

A single atom is an example of a ``physically unary'' quantum
computer, having a limited natural tensor-product structure provided
by the small number of physical degrees of freedom.  Similar poor
scaling will be seen in any implementation consisting of a single
particle, a single atom, or a single molecule consisting of a fixed
number of atoms.  The fungibility of Hilbert spaces means that one
can impose an artificial tensor-product structure on the Hilbert
space of these systems, equivalent to that of qubits, but this does
not obviate the need to provide the physical resources to generate
orthogonal quantum states.  Without a scalable tensor-product
structure corresponding to a division into physical degrees of
freedom, the action resources along one or more of the physical
co\"ordinate axes must blow up exponentially with problem size,
meaning that these systems are not suitable for scalable quantum
computation.

This should be contrasted with quantum computing using multiple
atoms, containing a physical tensor product structure, such as in an
ion trap \cite{Kielpinski}.  Quantum information is stored in two
sublevels  of each of the ion's ground states and manipulated with a
limited number of vibrational states.  A Hilbert space of 100 qubits
requires 100 ions in their ground states occupying 100 local
positions.  Neither the internal nor the external degrees of freedom
of the atoms require physical resources that grow exponentially in
order to accommodate a $2^N$-dimensional Hilbert space.

\section{Role of entanglement}
\label{sec:entanglement}

Entanglement is a distinctive feature of quantum mechanics.  It is
clearly a resource for such quantum information protocols as
teleportation, yet its role in quantum computation remains unclear.
Some claim it is the property that powers quantum
computation \cite{Ekert,Jozsa}, while others downplay its
significance \cite{Lloyd,Knight}.  The situation has been clarified
considerably by the recent work of Jozsa and
Linden \cite{JozsaLinden}, who showed that for a qubit quantum
computer---the extension to qudits is probably
straightforward---entanglement among all the qubits is a prerequisite
for an exponential speed-up over a classical computation.  The
Jozsa-Linden proof proceeds by showing that if entanglement extends
only to a fixed number of qubits, independent of problem size, the
computation can be simulated efficiently on a classical computer.

The Jozsa-Linden argument {\em assumes\/} a strictly scalable
tensor-product structure.  The global entanglement that accompanies
exponential speed-up is a consequence of assuming this tensor-product
structure and an initial pure state.  Consider a computation with an
exponential speed-up on a qubit quantum computer.  Mapped onto a
unary machine, the same computation produces {\em no\/} entanglement.
Whether run on the unary computer or the qubit computer, the
computation accesses arbitrary states---i.e., arbitrary
superpositions---in the computer's Hilbert space and has no efficient
description in the realistic language of classical computation.
Hilbert spaces are fungible!  Global entanglement is a result of
running the computation on a quantum computer with a tensor-product
structure; for such a computer, arbitrary superpositions lead to
entanglement among all the parts, because the states without such
global entanglement occupy only a tiny corner of Hilbert
space \cite{Ekert,Jozsa}.  On a physically unary computer, the same
arbitrary superpositions have no entanglement.

The global entanglement in a quantum computation is thus a
consequence of the need to save resources, which is what dictates a
strictly scalable tensor-product structure to start with. A weak
statement is that global entanglement is a measure of the computer's
ability to economize on physical resources.  A stronger conclusion is
that global entanglement is indeed the key quantum {\em resource\/}
that allows a scalable quantum computer to avoid an exponential
demand for physical resources.  For a pure-state quantum computer,
such as those considered here, entanglement is what allows the
arbitrary superpositions required for quantum computation to be
folded into a compact hypercube in phase space, as illustrated in
Fig.~\ref{fig1}.\footnote{In their analysis, Jozsa and Linden were
careful to point out that although entanglement among all qubits is
necessary for exponential speed-up, it is not sufficient: as shown by
Gottesman and Knill \cite{NielsenChuang}, there are sequences of
quantum gates that can be simulated efficiently on a classical
computer even though they entangle all qubits.  This does not
undermine the conclusion that entanglement is necessary for avoiding
an exponential resource demand, but it does suggest that some other
quantity, presumably closely related to entanglement, might
characterize more completely the ability of a quantum computer to
economize on physical resources.}

This line of reasoning, based on consideration of pure-state quantum
computation, is supported by what is known about mixed-state quantum
computation.  An example is provided by liquid-state nuclear magnetic
resonance (NMR) \cite{Cory,Jones}.  In NMR the qubits are the active
nuclear spins within each molecule in the liquid sample, and each
molecule is an independent quantum computer, which runs an
independent version of the computation.  Mixed states don't have any
effect on the analysis of this paper, which shows that to avoid an
exponential resource demand requires a scalable tensor-product
structure---in NMR this means that the number of qubits per molecule
must increase linearly, just as in a pure-state qubit computer---but
the further argument that entanglement follows from accessing
arbitrary states in a system with a tensor-product structure, does
not work for mixed states \cite{JozsaLinden}.  Indeed, with present
polarizations, the states accessed in NMR are known to be unentangled
up to about 23 qubits \cite{Braunstein,Menicucci,Gurvits} and, for
bigger numbers of qubits, are likely to be far less entangled than in
a corresponding pure-state quantum computer.

In a mixed-state quantum computer, the previous connection between
physical resources and entanglement is severed, but a new connection
arises to take its place.  The paucity of entanglement in mixed-state
quantum computing betrays another resource problem.  This resource
problem comes about from the need to read out the output reliably in
the presence of the noise produced by the mixed state and results in
a demand for a number of repetitions of the computation that
increases exponentially with problem size.  In NMR this resource
problem shows up as a demand for an exponentially increasing number
of molecules.  This mixed-state resource problem, which is apparently
independent of the phase-space considerations that we apply to
quantum computers in this paper, provides further evidence for the
importance of entanglement in avoiding exponential resource demands
in quantum computation.

\section{Conclusion}
\label{sec:conclusion}

Our contention in this paper is that the fundamental requirement for
a scalable quantum computer is set by the need to economize on
physical resources in providing the primary resource of Hilbert-space
dimension.  To avoid an exponential demand for physical resources,
the number of degrees of freedom must grow quasilinearly with the
equivalent number of qubits.  This requirement means that a scalable
quantum computer must have a robust tensor-product structure.
Systems without such a tensor-product structure are not suitable for
scalable quantum computation.

Physical systems that don't scale properly, such as liquid-state NMR,
Rydberg atoms, or molecular magnets, are still worth studying for a
variety of reasons.  First and foremost, they embody fundamental
physical questions that are worth investigating in their own right,
regardless of their relevance to quantum information science. Second,
they can be used to develop new technologies for control, readout,
and error correction in quantum systems.  These new technologies
might have applications to quantum-information-processing jobs
outside quantum computation, and they might be transferable to
scalable quantum computers.  Finally, the scalability criteria
formulated in this paper are asymptotic requirements.  They are
useful for assessing the physical resources required for a
general-purpose quantum computer to do problems of increasing size.
Yet even for this purpose, they are imperfect tools, because no real
computer is expected to do problems of arbitrary size.  Nonscalable
systems might be able to provide sufficient Hilbert-space dimension
for special-purpose quantum computations that need only a limited
number of qubits, such as simulation of other quantum
systems \cite{Lloydsim}.

Hilbert space is essential for quantum computation.  Yet it is an odd
sort of thing to need.  It is not a physical object, but rather a
mathematical abstraction in which we describe physical objects.  A
Hilbert space gets a physical interpretation---a connection to the
external world---only through the physical system that we describe in
that Hilbert space.  The connection is made through privileged
observables---the generators of space-time symmetries, e.g.,
position, momentum, angular momentum, and energy---which determine a
set of physical degrees of freedom for the system.  This connection
made, we can determine how the physical resources, measured in terms
of phase-space actions constructed from the privileged observables,
must grow in order to provide the Hilbert-space dimension needed for
a quantum computation.

Our degrees-of-freedom analysis can be applied to the physical
resources required by a classical computer.  Generally the subsystems
in a classical computer consist of many physical degrees of freedom.
If each distinguishable configuration of a subsystem occupies a fixed
phase-space volume, then our analysis shows that the physical
resources required by the classical computer grow exponentially
unless the number of subsystems grows quasipolynomially with problem
size.  But the phase-space scale in the classical analysis is not
fundamental, instead being set by noise and the resolution of
measuring devices.  This makes the classical analysis of resource
requirements dependent on other features of a classical computer. The
difference for a quantum computer is that Planck's constant sets a
fundamental scale, which makes the resource requirements presented
here prerequisites for scalable quantum computation, prior to the
other necessary requirements for a quantum computer's operation.

\section*{Acknowledgments}
This work was partly supported by the National Security Agency (NSA)
and the Advanced Research and Development Activity (ARDA) under Army
Research Office (ARO) Contract No.~DAAD19-01-1-0648 and by the Office
of Naval Research under Contract Nos.~N00014-00-1-01578 and
N00014-99-1-0247.  This work grew in large part out of discussions at
the Institute Theoretical Physics of the University of California,
Santa Barbara, where the authors were in residence during the fall of
2001.  The authors received support from the ITP's National Science
Foundation Contract No.~PHY99-07949.

\end{document}